\documentclass[sigconf]{acmart}
\AtBeginDocument{%
  }

\copyrightyear{2026}
\acmYear{2026}
\setcopyright{cc}
\setcctype{by}
\acmConference[ISWC '26]{Proceedings of the 2026 ACM International Symposium on Wearable Computers}{October 11--15, 2026}{Shanghai, China}
\acmBooktitle{Proceedings of the 2026 ACM International Symposium on Wearable Computers (ISWC '26), October 11--15, 2026, Shanghai, China}
\acmDOI{10.1145/3830727.3834835}
\acmISBN{979-8-4007-2872-3/2026/10}

\usepackage{enumitem}
\usepackage{multirow}
\usepackage{xspace}
\usepackage{color, colortbl}
\newcommand{\blead}[1]{\indent\textbf{#1}} 

\newcommand{\eg}{\textit{e.g.}\@\xspace}

\newcommand\revised[1]{\textcolor{black}{#1}} 
\definecolor{lightgray}{gray}{0.9}

\def\MlpLosoYoung{75.8\%}

\def\LimuBertLosoYoung{81.5\%}

\def\MlpLosoOlder{34.4\%}

\def\AttnTcnLosoOlder{57.1\%}

\def\HcMlpLosoOlder{56.1\%}
\def\ResnetLosoOlder{40.0\%}
\def\ResnetFrozenLosoOlder{38.9\%}
\def\SSLFrozenLosoOlder{69.5\%}

\def\MlpLosoGap{41.4\%}

\def\TransformerLosoGap{44.6\%}
\def\AttnTcnLosoGap{23.9\%}

\def\SSLFrozenLosoGap{12.1\%}

\def\AvgLosoRwOlderGap{21.2\%}

\def\SSLFrozenCdYoung{84.4\%}

\def\MlpCdOlder{38.4\%}

\def\MobileHartCdOlder{60.5\%}

\def\SSLFrozenCdOlder{65.1\%}

\def\MlpCdGap{32.4\%}

\def\SSLFrozenCdGap{19.3\%}

\begin{document}

\title[Characterizing the Performance Gap]{Characterizing the Performance Gap in Human Activity Recognition for Older Adults}

\author{Hossein Khayami}
\email{khayami@umd.edu}
\orcid{0000-0002-4485-9322}
\affiliation{%
  \institution{University of Maryland}
  \city{College Park}
  \state{Maryland}
  \country{USA}
}

\author{Sungjin Hwang}
\email{sungjin@umd.edu}
\orcid{0000-0003-2685-5544}
\affiliation{%
  \institution{University of Maryland}
  \city{College Park}
  \state{Maryland}
  \country{USA}
}

\author{Eshed Ohn-Bar}
\email{eohnbar@bu.edu}
\orcid{0009-0009-1607-4315}
\affiliation{%
  \institution{Boston University}
  \city{Boston}
  \state{Massachusetts}
  \country{USA}
}

\author{David E. Conroy}
\email{deconroy@umich.edu}
\orcid{0000-0003-0204-4093}
\affiliation{%
  \institution{University of Michigan}
  \city{Ann Arbor}
  \state{Michigan}
  \country{USA}
}

\author{Amanda Lazar}
\email{lazar@umd.edu}
\orcid{0000-0001-8110-4239}
\affiliation{%
  \institution{University of Maryland}
  \city{College Park}
  \state{Maryland}
  \country{USA}
}

\author{Eun Kyoung Choe}
\email{choe@umd.edu}
\orcid{0000-0001-5038-8320}
\affiliation{%
  \institution{University of Maryland}
  \city{College Park}
  \state{Maryland}
  \country{USA}
}

\author{Hernisa Kacorri}
\email{hernisa@umd.edu}
\orcid{0000-0002-7798-308X}
\affiliation{%
  \institution{University of Maryland}
  \city{College Park}
  \state{Maryland}
  \country{USA}
}
\renewcommand{\shortauthors}{Hossein Khayami et al.}

\begin{abstract}
Human activity recognition (HAR) from wrist-worn accelerometers is increasingly used for health and behavioral tracking. Yet, most wearable HAR models are developed and evaluated on datasets dominated by younger adults, leaving it unclear whether benchmark progress generalizes across age groups. In this work, we leverage \revised{MyMove, our carefully annotated, free-living} older-adult HAR dataset (mean age 71), to evaluate deep-learning architectures and training regimes under both leave-one-subject-out and cross-dataset transfer.
We find that improvements on younger-adult benchmarks fail to transfer equally to \revised{data collected from} older adults, \revised{resulting in} a persistent and often widening performance gap.
However, richer representations, particularly frozen self-supervised features pretrained on the age-diverse UK Biobank dataset, substantially improve performance \revised{on data from older adults} and consistently narrow the performance gap, at modest cost to younger-adult performance, though disparities remain. 
These findings suggest that benchmark gains and architectural scaling alone provide an incomplete picture of progress in wearable HAR, and broader advances may \revised{require} representations that \revised{better} capture population diversity, alongside personalized adaptation to individual movement patterns and routines.

\end{abstract}





\maketitle

\section{Introduction} \label{sec:intro}
Despite substantial advances, human activity recognition \revised{(HAR) using} wrist-worn accelerometer \revised{data remains a challenging problem}. Wrist motion \revised{alone provides an incomplete and often ambiguous view of human activity,} as similar wrist movements can arise from different postures and activities~\cite{pannurat2017analysis}. \revised{Moreover, everyday behavior rarely conforms to discrete activity categories:} activities such as grocery shopping or household chores interleave standing, walking, and transitional movements that resist clean categorical labels. \revised{Beyond these intrinsic challenges, progress is difficult to compare as studies differ in} sensing setups, activity taxonomies, ecological conditions, and evaluation protocols. 
\revised{Recent advances in deep learning have nevertheless led to} steady improvements on widely used HAR datasets. Yet most publicly used collections are built predominantly from younger adults performing researcher-scripted activities under controlled conditions~\cite{reiss2012introducing, roggen2010collecting, mhealth_319, weiss2019wisdm, sztyler2016body}. 
\revised{Consequently, models are typically trained and evaluated on data representing relatively narrow movement distributions, while datasets capturing free-living behavior from older adults remain rare.}
This mismatch matters because wearable \revised{activity trackers} are increasingly used across broad adult populations to track physical activities~\cite{brickwood2019consumerbased}, infer health and well-being~\cite{chandrasekaran2020patterns}, and promote behavioral change through feedback~\cite{feng2021self, evenson2015systematic}. However, sustained use among older adults is often limited, with perceived inaccuracies in device outputs cited as one of the main reasons for discontinuation~\cite{AARP2016Building}.
Older and younger adults may perform the same activity with different rhythms, assistive devices, routines, and movement patterns~\cite{hultsch_variability_2002, kang_separating_2008, callisaya_ageing_2010}, yet current evaluation practices provide little visibility into whether architectural improvements generalize across these populations. 

To examine whether benchmark progress \revised{generalizes to data from older adults}, we evaluate six widely used HAR architectures under both within-dataset leave-one-subject-out (LOSO) and cross-dataset transfer. \revised{Our primary} comparison contrasts younger adults from the RealWorld dataset~\cite{sztyler2016body} with older adults from \revised{MyMove,} our free-living smartwatch dataset, while retaining \revised{the age outlier participants in each dataset to probe potential dataset-specific effects}.
\revised{To examine} the role of representation, we fix the classifier head while varying the feature extractor, \revised{comparing raw inputs, randomly initialized models, and supervised representations with} handcrafted features and self-supervised representations, including a backbone pretrained on the age-diverse UK Biobank dataset~\cite{doherty2017large, yuan2024self}. 

Across multiple modern HAR architectures, we find that improvements on younger-adult benchmarks fail to transfer proportionally to data from older adults, producing a persistent and sometimes widening performance gap. Under LOSO evaluation, the gap ranges from \AttnTcnLosoGap{} to \TransformerLosoGap{} across architectures, and top-ranked younger-adult models do not reliably rank among the best for older adults. 
This pattern persist under both within-dataset LOSO and cross-dataset transfer.
A different picture emerges when the focus shifts from benchmark-ranked architectures to representation quality. Richer representations, particularly frozen self-supervised ones pretrained on age-diverse data, substantially improve performance on older-adults, narrowing the LOSO gap from \MlpLosoGap{} to \SSLFrozenLosoGap{} and the zero-shot gap from \MlpCdGap{} to \SSLFrozenCdGap{} at only modest costs to younger-adult performance (2\% and 4\%). This suggests that optimizing exclusively for the dominant benchmark population may not yield models that generalize robustly to older-adult data. 

This paper makes three contributions:

\begin{itemize}[leftmargin=*,nosep]
    \item \textit{Empirically}, we show that gains on dominant younger-adult HAR benchmarks do not translate to proportional improvements on data from older adults and can even widen the performance gap.

    \item \textit{Methodologically}, 
    rather than comparing architectures by chronology, we rank them by younger-adult benchmark performance, treating benchmark success as the field's operational notion of progress, and test whether it transfers to older-adult data.
    
    \item \textit{Conceptually}, our findings shift attention from architecture-level benchmark rankings toward representation quality, showing that age-diverse self-supervised pretraining improves performance on older-adult data while narrowing the performance gap.

\end{itemize} 
\section{Related Work}
\label{sec:related}

To situate our approach, we examine prior work on population representation in HAR datasets, neural architectures for wearable sensing, and evaluation practices for cross-dataset generalization.

\subsection{HAR Datasets \& Population Representation}
\label{sec:related:datasets}

Sensor-based HAR datasets have been collected predominantly from younger, healthy adults in controlled or semi-controlled settings~\cite{janidarmian2017comprehensive, kwapisz2011activity, sztyler2016body, ashry2020charm}. Classic benchmarks such as PAMAP2~\cite{reiss2012introducing} and MHealth~\cite{mhealth_319} exemplify the field's data collection paradigm: participants perform prescribed activities under supervision, maximizing label fidelity while limiting ecological realism and the diversity of natural movements~\cite{janidarmian2017comprehensive}. Large-scale free-living datasets remain rare. CAPTURE-24 (151 participants, over 3,800 hours of wrist-worn data) and the UK Biobank ($\sim$100,000 participants spanning a broad adult age range) are exceptions in scale and ecological validity~\cite{chan2024capture24, yuan2024self}, yet, CAPTURE-24's oldest group is 53+, and the UK Biobank has no activity labels. Datasets centered on older adults are substantially less common~\cite{garcia-moreno2022machine}. ADAPT (aged 68--90, free-living home protocol with wrist-worn sensors)~\cite{bourke2017physical} and HAR70+ (aged 70--95)~\cite{ustad2023validation} are notable efforts. However, ADAPT relied on a dense 12-sensor setup that limits ecological realism, and HAR70+ used thigh and lower-back sensors rather than wrist-worn devices (\eg, smartwatches) that reflect everyday adoption. As a result, free-living, wrist-worn HAR datasets with reliable activity labels and substantial older-adult representation remain extremely limited~\cite{khayami2025verbal}. This imbalance matters for evaluation as models are typically optimized and compared within homogeneous younger-adult populations, leaving age-related generalization gaps underexplored.

\subsection{Evolution of Neural Architectures for HAR}
\label{sec:related:architectures}

Wearable HAR has repeatedly adopted neural architectures from other machine-learning domains, including convolutional networks, recurrent sequence models, Transformers, and self-supervised pretraining~\cite{krizhevsky2012imagenet, hochreiter1997long, vaswani2017attention, devlin2019bert, chen2020simclr}. These were later adapted to wearable sensing via CNN-based activity recognition~\cite{yang2015deep}, LSTMs for body-worn streams~\cite{chen2016lstm, ordonez2016deep, guan2017ensembles}, Transformer-based IMU encoders~\cite{mahmud2020human, xu2021limubert}, and large-scale self-supervised learning on accelerometer data~\cite{saeed2019multitask, yuan2024self}, with reported performance steadily improving. Yet this progress in HAR is hard to interpret chronologically. Models are often imported into wearable sensing years after their original development, and performance depends heavily on training procedures, datasets, and evaluation protocols. As noted, those evaluation datasets are \revised{drawn predominantly from younger-adult populations and are} relatively homogeneous. Consequently, improved performance may reflect better modeling of younger-adult movement without necessarily improving robustness for older adults. We therefore order architectures not by publication date but by their performance on younger-adult data, treating this ranking as a proxy for the field’s operational notion of progress, and ask whether those gains translate proportionally to older-adult data.

\subsection{Generalization \& Evaluation Across Datasets}
\label{sec:related:evaluation}

\revised{A central question for HAR is whether models transfer beyond the dataset they were trained on.} Recent work has increasingly examined cross-dataset generalization~\cite{napoli2024benchmark, hong2024crosshar, miao2024goat, lu2022sdmix} and most studies report a substantial source-to-target drop while treating the domain gap as a single aggregate phenomenon~\cite{zhou2024crossdomain, li2024crossdomain}. Where individual sources of shift are studied, emphasis falls on device heterogeneity~\cite{napoli2024benchmark}, sensor placement~\cite{thukral2023crossdomain}, or lab-to-free-living transitions~\cite{vaizman2016recognizing, bulling2014tutorial}; population shift, especially age-related transfer, is underexplored~\cite{pan2024multichannel}, and labeling-method effects are rarely isolated~\cite{khayami2025verbal}; analyses that jointly decompose these axes remain rare~\cite{dhekane2025transfer}. As a result, models trained on younger-adult, scripted data face simultaneous population, environment, and label shifts when deployed on free-living older adults. Existing protocols give limited visibility into how architectural improvements interact with these demographic shifts. We examine this gap by comparing multiple HAR architectures across younger- and older-adult datasets under both in-distribution and cross-dataset settings, \revised{asking whether benchmark gains generalize to data from older adults}.
\section{Experimental Setup}
\label{sec:setup}
We evaluate whether improvements in benchmark HAR performance \revised{generalize to data from older adults by} comparing multiple architectures under within- and cross-dataset protocols spanning younger- and older-adult populations.

\subsection{Datasets}
\label{sec:setup:datasets}

\begin{table}[t]
    \centering
    \caption{Smartwatch-based human activity recognition datasets used in the study. ${}^1$The number of activities reported here is after excluding uncommon activities from datasets. 
    }
    \label{tab:datasets}
    \small
    \vspace{-0.3cm}
    \resizebox{\columnwidth}{!}{%
    \begin{tabular}{lccccc}
        \toprule
        \textbf{Dataset} & \textbf{Pop.} & \textbf{Age}   & \textbf{\#Part.} &  \textbf{\#Act.${}^1$}  & \textbf{\#Win}  \\ 
        \midrule
        
        {RealWorld (RWy)}~\cite{sztyler2016body} & younger &  29.7 (16--52) & 14 & 5 & 8{,}900\\
        \rowcolor{gray!15}{RealWorld (RWo)}~\cite{sztyler2016body} & older & 62& 1 & 5 & 736\\
        \revised{MyMove (MMo)} & older & 71.1 (61--90) & 13 & 5 & 17{,}160\\

        \rowcolor{gray!15}\revised{MyMove (MMy)} & \revised{younger} & \revised{25.5 (20--31)} & \revised{2} & \revised{3} & \revised{418}\\
         \midrule
        
        {WISDM}~\cite{weiss2019wisdm} & students & -- & 46 & 4 & 6{,}230 \\
        \rowcolor{gray!15}{UK Biobank}~\cite{doherty2017large} & broader  & 62 (45–79) & 103k & -- & $\sim$1B \\

         \bottomrule
    \end{tabular}
    }
    \vspace{-0.42cm}
\end{table}

Table~\ref{tab:datasets} summarizes the four datasets and their roles.  WISDM~\cite{kwapisz2011activity} and RealWorld~\cite{sztyler2016body} are widely used wrist-worn HAR benchmarks of younger-adult, researcher-supervised data collection.  
Our \revised{MyMove} dataset\revised{~\cite{kim2022mymove,khayami2025verbal}} provides free-living older-adult recordings and is our primary evaluation target.
UK Biobank~\cite{doherty2017large} is used only for self-supervised representation learning, given its scale and demographic diversity. All signals are preprocessed to a common format: triaxial accelerometer windows resampled to 30~Hz and segmented into non-overlapping 10-second windows of 300 samples, in units of $g$ (gravity included). We adopt a 5-class activity taxonomy: \textit{Lying}, \textit{Sitting}, \textit{Standing}, \textit{Walking}, and \textit{Running}; chosen to maximize activity overlap across datasets.

\blead{WISDM}~\cite{kwapisz2011activity} contains scripted 
LG smartwatch recordings from 51 participants in an indoor lab. Five were excluded for missing at least one class. Of its 18 classes; we retain four matching our taxonomy (\textit{Sitting}, \textit{Standing}, \textit{Walking}, and \textit{Running}); \textit{Lying} is absent.

\blead{RealWorld}~\cite{sztyler2016body} contains LG/Samsung smartwatch recordings of 15 participants performing five outdoor activities, 14 of them younger adults (RWy). Subject~5, aged 62, is the only older \revised{participant (RWo), retained as a qualitative reference to probe potential dataset-specific effects.}

\begin{figure}
    \centering
       \includegraphics[width=1\linewidth]{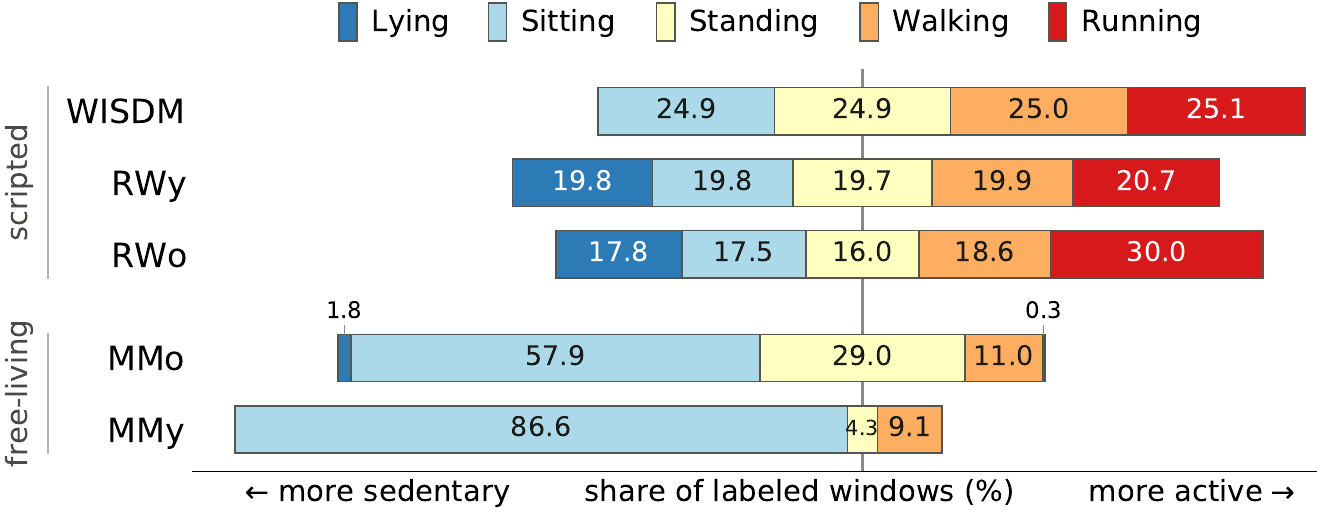}
     \vspace{-0.7cm}
    \caption{\revised{Activity composition across datasets; each row sums to 100\% of that population's labeled windows.}}
    \label{fig:activity_dist}
    \Description[Horizontal stacked bars of activity-class proportions in each dataset.]{Horizontal stacked bar chart comparing activity-class proportions across five datasets, grouped into scripted (WISDM, RealWorld younger, RealWorld older) and free-living (MyMove older, MyMove younger). Each bar spans the full width and is divided into five segments for Lying, Sitting, Standing, Walking, and Running. The scripted datasets are close to balanced: WISDM is roughly a quarter each of Sitting, Standing, Walking, and Running with no Lying; RealWorld younger is near 20 percent for every class; RealWorld older is similar but 30 percent Running. The free-living datasets are heavily skewed toward stationary activity: MyMove older is 57.9 percent Sitting, 29.0 Standing, 11.0 Walking, 1.8 Lying, and 0.3 Running; MyMove younger is 86.6 percent Sitting, 4.3 Standing, and 9.1 Walking, with no Lying or Running.}
    \vspace{-0.4cm}
\end{figure}

\blead{\revised{MyMove}}\revised{~\cite{kim2022mymove}} contains free-living smartwatch recordings from 13 older adults \revised{(MMo)}, ages 61--90, \revised{and two younger adults (MMy), ages 20--31,}  each wearing a Fossil~Gen~5 smartwatch during wake time for seven consecutive days in daily life. \footnote{\revised{Additional details, visualizations, and related publications describing the MyMove dataset are available at \url{https://mymove-collective.github.io/dataset/}. We are working toward a public release of a privacy-preserving subset of the dataset, which will be made available through this project page.}}
Activity labels were obtained through consensus triangulation combining predictions from a thigh-worn sensor and participant verbal reports\revised{~\cite{khayami2025verbal}}. The resulting class distribution shown in Figure~\ref{fig:activity_dist} reflects genuine free-living behavior. \revised{Because the MMy data do not contain instances of \textit{Lying} and \textit{Running}, they are excluded from the quantitative evaluation but used for qualitative interpretation where appropriate.}

\blead{UK Biobank}~\cite{doherty2017large} provides over 700,000 person-days of wrist accelerometer recordings from more than 100,000 adults, including 46,605 participants (45\%) aged 65 or older. Although the dataset contains no activity labels, its scale and demographic diversity makes it well suited for representation learning. The dataset is no longer publicly available; we use the self-supervised learning (SSL) ResNet backbone pretrained by Yuan et al. ~\cite{yuan2024self} on this dataset in our SSL-based configurations (Section~\ref{sec:setup:arch}).

\subsection{Architectures}
\label{sec:setup:arch}
We evaluate architectures at two levels: end-to-end HAR models and representation schemes. The first \revised{tests whether the performance gap between younger- and older-adult data} persists; the second examines the contribution of the learned representation.

The \textbf{end-to-end models} comprise six architectures spanning major paradigms in wearable HAR, from feedforward baselines to modern attention-based models: a minimal MLP baseline, the widely adopted DeepConvLSTM~\cite{ordonez2016deep}, the efficiency-oriented MobileHART~\cite{ek2023transformerbased}, and three more recent attention-based architectures: AttnTCN~\cite{wei2024tcnattentionhar}, Transformer~\cite{vaswani2017attention}, and LIMU-BERT~\cite{xu2021limubert}. Together they span increasing complexity and temporal-modeling capacity.

The \textbf{representation schemes} focus on the representation layer by fixing a common MLP classifier head and  varying only the input representation. We evaluate six configurations spanning increasing representational structure, from raw inputs and random priors to human-designed features and learned supervised and self-supervised representations. These configurations are: raw accelerometer windows passed directly to an MLP classifier, a randomly initialized frozen ResNet~\cite{he2016deep}, a fully fine-tuned ResNet from random initialization, 32-dimensional handcrafted features (Table~\ref{tab:handcrafted}), a fully fine-tuned ResNet-SSL, and a frozen ResNet-SSL backbone. Even frozen, randomly initialized networks retain useful structure~\cite{frankle2021training}. This setup \revised{provides a controlled progression or representation quality for comparison} (Section~\ref{sec:results:representation}).

\begin{table}[t]
  \centering
  \caption{Thirty-two handcrafted accelerometry features typically appearing in prior HAR literature \eg, \cite{yuan2024self, narayanswamy2024scaling}.}
  \label{tab:handcrafted}
  \small
  \vspace{-0.35cm}
  \footnotesize
  \setlength{\tabcolsep}{4pt}
  \resizebox{\columnwidth}{!}{%
  \begin{tabular}{llc}
    \toprule
    \textbf{Feature} & \textbf{Description} & \textbf{Source} \\
    \midrule
    mean\_\{x,y,z\}  & Per-axis mean                        &\cite{yuan2024self} \\
    std\_\{x,y,z\}   & Per-axis standard deviation          &\cite{yuan2024self} \\
    range\_\{x,y,z\} & Per-axis range (max $-$ min)         &\cite{yuan2024self} \\
    \midrule
    \revised{mag\_\{mean,std,range\}}     & Vector-magnitude mean, standard deviation, range                   &\cite{yuan2024self}\\
    mag\_mad      & Median absolute deviation of magnitude  &\cite{yuan2024self}\\
    \revised{mag\_\{kurtosis,skew\}} & Kurtosis and skewness of magnitude                   & \cite{yuan2024self,narayanswamy2024scaling} \\
    \midrule
    corr\_\{xy,xz,yz\} & Pearson correlation, across axis pairs       &\cite{yuan2024self} \\
    \midrule
    \revised{f1, f2}  & First and second dominant frequency in 0.5--6\,Hz          &\cite{yuan2024self}    \\
    ber & Band energy ratio: $E_{2\text{--}4\,\text{Hz}} / E_{0.5\text{--}2\,\text{Hz}}$ & this \\
    hr  & Harmonic ratio: $S(2f_1)/S(f_1)$              & this \\
    \midrule
    rms        & Root-mean-square of vector magnitude   & this \\
    log\_energy & Log of summed per-axis RMS             & \cite{narayanswamy2024scaling} \\
    \midrule
    log\_energy\_ratio & $\log(E_\text{PC1}/E_\text{RMS})$: energy on dominant axis & \cite{narayanswamy2024scaling} \\
    cov\_condition     & $\lambda_\text{max}/\lambda_\text{min}$: motion anisotropy & \cite{narayanswamy2024scaling} \\
    jerk\_auto\_ratio  & Lag-1 autocorrelation of PC1 / $E_\text{PC1}$: smoothness & \cite{narayanswamy2024scaling} \\
    \midrule
    \revised{zc\_\{avg, std\}} & Mean and Std. of inter-zero-crossing interval of PC1 & \cite{narayanswamy2024scaling} \\
    \midrule
   arm\_tilt  & Log amplitude in the sagittal ($x$--$z$) plane & \cite{narayanswamy2024scaling} \\
    \midrule
    step\_count & Peak count in magnitude above mean, $\geq$0.3\,s apart & \cite{narayanswamy2024scaling} \\
    sleep\_coef & Per-axis amplitude sum digitized into log-scaled bins  & \cite{narayanswamy2024scaling} \\
    \bottomrule
  \end{tabular}
  }
       \vspace{-0.5cm}
\end{table}

\subsection{Evaluation Protocol}
\label{sec:setup:protocol}

\indent\blead{Training.} All models use the Adam optimizer with learning rate~$10^{-3}$ and cross-entropy loss with inverse-frequency class weights, trained for 30 epochs with no fine-tuning on any evaluation dataset.

\blead{Evaluation.} We evaluate all architectures under two protocols. \emph{Leave-one-subject-out (LOSO)} trains on all but one participant and evaluates on the held-out one, applied independently within RealWorld (younger adults, $n{=}14$) and \revised{MyMove} (older adults, $n{=}13$). \emph{Cross-dataset (zero-shot)} trains on WISDM and evaluates directly on both populations. Applying both to the same architectures tests whether the age gap is an artifact of the evaluation setting or a deeper limitation of current HAR models.

\blead{Metric.} We report pooled macro F1: windows are pooled across participants and F1 is macro-averaged, giving each activity class equal weight. \revised{As expected for free-living data,} our older-adult dataset is strongly class-imbalanced: \textit{Sitting} and \textit{Standing} account for 87\% of windows, while \textit{Lying} and \textit{Running} each comprise less than 2\%. Weighted F1 would downweight these rare but behaviorally important classes in proportion to their frequency; macro F1 weights them equally.

\section{Results}
\label{sec:results}
Across in-distribution and cross-dataset evaluations, improvements \revised{in HAR  performance on RealWorld younger adults do not translate proportionally to MyMove older adults, indicating a persistent performance} disparity.
However, this gap progressively narrows as representation quality improves through self-supervised pretraining on age-diverse data.

\subsection{Benchmark Gains Do Not Close the Gap}
\label{sec:results:gap}

\begin{figure}[t]
  \centering
  \includegraphics[width=\columnwidth]{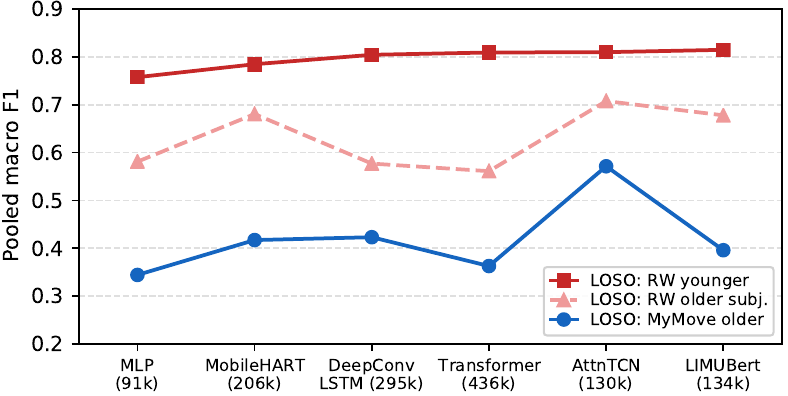}
           \vspace{-0.7cm}
  \caption{Benchmark gains on younger adults fail to transfer equally to older adults. Architectures are sorted by RWy performance; parameter counts in parentheses.}
  \label{fig:loso_rw_young}
   \Description[Line chart of leave-one-subject-out macro F1 for six architectures, by population.]{Line chart with three series over six deep-learning architectures on the horizontal axis, ordered left to right by ascending RealWorld younger-adult performance: MLP (91k parameters), MobileHART (206k), DeepConvLSTM (295k), Transformer (436k), AttnTCN (130k), and LIMUBert (134k). The vertical axis is pooled macro F1 from 0.2 to 0.9. The RealWorld younger series stays highest, rising gently from about 0.76 to 0.81. The RealWorld older single-subject series is in the middle, varying between about 0.56 and 0.70. The MyMove older series stays lowest, between about 0.34 and 0.57, and does not follow the left-to-right ordering: it peaks at AttnTCN rather than at the architectures that are best for younger adults. The wide and roughly constant gap between the younger and MyMove older series is the key point: gains along the younger-adult ordering do not reach older adults.}
      \vspace{-0.5cm}
\end{figure}

Figure~\ref{fig:loso_rw_young} shows LOSO performance for six deep-learning architectures on younger adults (RealWorld, $n{=}14$) and older adults (\revised{MyMove}, $n{=}13$), sorted by ascending younger-adult benchmark performance. As architectures achieve incremental gains on the younger benchmark, those gains do not translate proportionally to \revised{MyMove} older-adult data. The younger-adult line rises smoothly from \MlpLosoYoung~ to \LimuBertLosoYoung, while the older-adult line fluctuates between \MlpLosoOlder~ and \AttnTcnLosoOlder;  the architectures that rank highest on younger adults are not those that rank highest on older adults. RealWorld subject~5, the single older participant within RealWorld, broadly tracks the \revised{MyMove} older-adult curve (\AvgLosoRwOlderGap~ higher on average), while the gap between RealWorld younger adults and \revised{MyMove} older adults is large and fluctuating, from \AttnTcnLosoGap~ for AttnTCN to \TransformerLosoGap~ for Transformer. This suggests that \revised{differences associated with participant age contribute to the observed gap, alongside} dataset-specific factors such as labeling approach and class imbalance. 
\revised{The aggregate gap also masks uneven per-class performance. For AttnTCN, the architecture with the smallest overall gap, F1 on MyMove older adults ranged from 13\% for \textit{Lying} to 81\% for \textit{Sitting}, whereas the same model scored a more uniform 69–97\% across all classes on RealWorld younger adults (Table~\ref{tab:attntcn_perclass_f1}).}
Incremental gains on standard younger-adult benchmarks may therefore provide an incomplete, and potentially misleading, picture of progress in HAR. 
Architectures that appear increasingly effective on dominant benchmarks may still fail to reduce, or even widen, performance disparities. 

\subsection{The  Gap Persists Across Protocols}
\label{sec:results:structural}

\begin{figure}[t]
  \centering
  \includegraphics[width=\columnwidth]{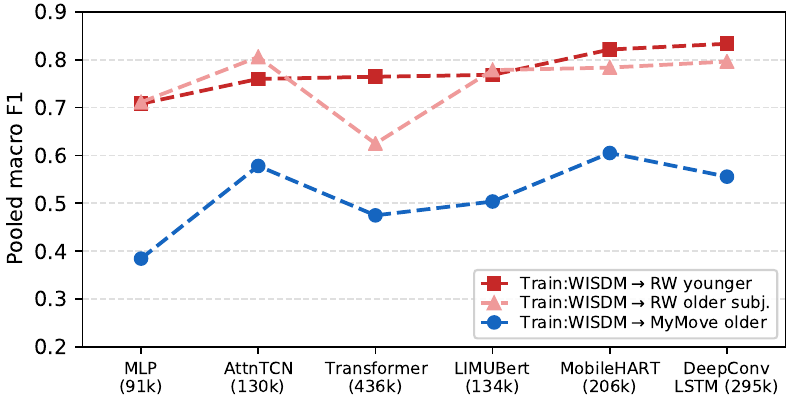}
    \vspace{-.7cm}
  \caption{Zero-shot benchmark gains on younger adults fail to transfer equally to older adults. Architectures are sorted by WISDM$\to$RWy performance.}
  \label{fig:cd_rw_young}
  \Description[Line chart of zero-shot cross-dataset macro F1 for six architectures, by population.]{Line chart of zero-shot cross-dataset performance, with models trained on WISDM and evaluated without fine-tuning. Six architectures on the horizontal axis are ordered left to right by ascending WISDM-to-RealWorld younger performance: MLP (91k parameters), AttnTCN (130k), Transformer (436k), LIMUBert (134k), MobileHART (206k), and DeepConvLSTM (295k). The vertical axis is pooled macro F1 from 0.2 to 0.9, and all three series are dashed. The RealWorld younger series rises from about 0.71 to 0.83. The RealWorld older single-subject series tracks close to it, dipping to about 0.62 at Transformer. The MyMove older series stays well below, between about 0.39 and 0.60. As in the leave-one-subject-out figure, the MyMove older series remains far under the younger series across all architectures, so benchmark gains again fail to transfer to older adults under cross-dataset evaluation.}
    \vspace{-0.4cm}
\end{figure}

Figure~\ref{fig:cd_rw_young} repeats the analysis under cross-dataset evaluation: WISDM-trained models are evaluated zero-shot on younger adults (RealWorld) and older adults (\revised{MyMove older} and RealWorld subject~5), sorted by ascending RealWorld younger performance. Despite the shift from in-distribution LOSO to zero-shot transfer, the same pattern persists. \revised{MyMove} older-adult performance remains fluctuating and consistently lower, ranging from \MlpCdOlder~ for MLP to \MobileHartCdOlder~ for MobileHART, and the gap does not close along the benchmark axis; if anything it widens. 
\revised{Under zero-shot transfer, per-class F1 for AttnTCN stayed low across all classes for MyMove older adults, from 45\% (\textit{Standing}/\textit{Walking}) to 73\% (\textit{Running}), well below the 94\% (\textit{Walking}) that younger adults reached (younger range 50--94\%; Table~\ref{tab:attntcn_perclass_f1}).}
This persistent gap under both evaluations, suggests that gains on dominant younger-adult benchmarks \revised{do not necessarily translate into robust cross-population performance}. The RealWorld older adult again tracks the \revised{MyMove older-adult pattern, providing qualitative evidence that the observed divergence is not solely attributable  to dataset-specific collection conditions}. The disconnect between younger- and older-adult performance \revised{persists across both within-dataset generalization and cross-dataset transfer, suggesting that it is not simply} an artifact of a particular evaluation strategy.

\begin{table}[t]
\centering
\caption{\revised{Per-class F1 for AttnTCN under LOSO and cross-dataset (WISDM$\to$RealWorld/MyMove) evaluation.}}
\label{tab:attntcn_perclass_f1}
\small
\vspace{-.25cm}
\begin{tabular}{llccccc}
\toprule
Evaluation & Group & \textit{Lie} & \textit{Sit} & \textit{Stand} & \textit{Walk} & \textit{Run} \\
\midrule
\multirow{2}{*}{LOSO} & MMo & 13\% & 81\% & 58\% & 64\% & 70\% \\
 & RWy & 70\% & 69\% & 80\% & 97\% & 88\% \\
\midrule
\multirow{2}{*}{Cross-dataset} & MMo & -- & 67\% & 45\% & 45\% & 73\% \\
& RWy & -- & 50\% & 67\% & 94\% & 93\% \\
\bottomrule
\end{tabular}
\vspace{-.7cm}
\end{table}

\begin{figure*}[t]
  \centering
  \vspace{-0.1cm}
  \includegraphics[width=\textwidth]{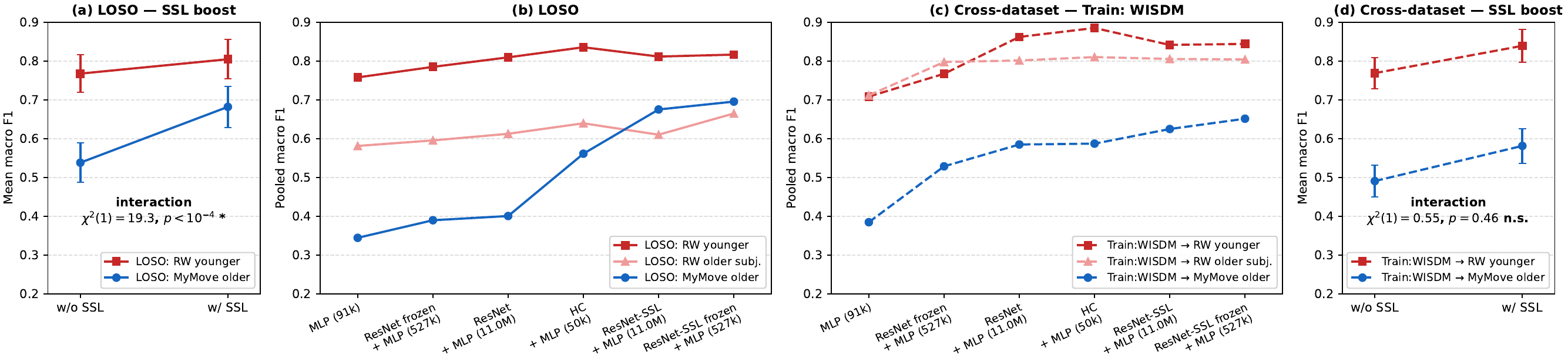}
        \vspace{-0.6cm}
  \caption{Increasingly structured representations produce more aligned performance between younger and older adults, with the frozen SSL-pretrained ResNet backbone strongest. \revised{(a) LOSO mean macro F1 (95\% CIs) for RealWorld younger and MyMove older adults, without vs.\ with SSL.} (b,c) LOSO and cross-dataset (Train:WISDM) pooled macro F1 by population across the six configurations; \revised{(d) similar to~(a), for cross-dataset.}}
  \label{fig:mlp_head_loso_cd}
  \Description[Four panels showing structured representations narrow the younger-older performance gap.]{Four panels sharing a vertical axis of macro F1 from 0.2 to 0.9. Panels a and d are narrow interaction plots; panels b and c are wider line charts. In panel a (leave-one-subject-out), two lines connect performance without self-supervised learning to performance with it: RealWorld younger rises slightly from about 0.77 to 0.80, while MyMove older rises steeply from about 0.54 to 0.68, so the lines converge; an annotation reports a significant cohort-by-representation interaction, chi-squared 19.3 and p below ten to the minus four. Panels b and c plot pooled macro F1 across six MLP-head configurations ordered by increasing representation quality (standalone MLP, frozen random ResNet, trained ResNet, handcrafted features, ResNet-SSL, frozen ResNet-SSL) for RealWorld younger, a RealWorld older subject, and MyMove older; older-adult performance climbs steadily along this ordering under both leave-one-subject-out (b) and cross-dataset training on WISDM (c), with frozen ResNet-SSL highest. Panel d repeats the panel-a interaction for cross-dataset transfer, where both cohorts improve in parallel and the interaction is not significant (chi-squared 0.55, p equals 0.46). Richer representations lift older adults most under leave-one-subject-out, narrowing the age gap, but not under cross-dataset transfer.}
        \vspace{-0.3cm}
\end{figure*}

\subsection{Richer Representations Narrow the Gap}
\label{sec:results:representation}

Figure~\ref{fig:mlp_head_loso_cd} isolates the contribution of the representation by fixing the classifier to a single MLP head and varying only the feature extractor across six configurations under both LOSO (\revised{panels b}) and cross-dataset (\revised{panels c}) protocols\revised{; panels a and d collapse these into representations without vs.\ with SSL to test whether SSL narrows the gap}. These form a progression from raw input signals, to random weights, to handcrafted features, and finally to supervised and self-supervised representations. 
\revised{The x-axis orders these configurations according to increasing representational structure.}

Indeed, under LOSO, older-adult performance improves along the ladder. The standalone MLP, operating on raw windows without a representation layer, performs worst (\MlpLosoOlder). A randomly initialized frozen ResNet already helps (\ResnetFrozenLosoOlder): architectural inductive bias alone provides useful temporal structure even without learning. Letting its weights adapt to the training data improves further (\ResnetLosoOlder). Handcrafted features occupy a middle ground (\HcMlpLosoOlder): though not learned end-to-end, they encode interpretable hand-designed accelerometry descriptors such as cadence, spectral bands, and motion magnitude, surpassing a supervised ResNet while still below SSL versions. ResNet-SSL fully fine-tuned benefits from representations learned on the UK Biobank~\cite{yuan2024self}, capturing rich temporal patterns from large-scale free-living accelerometry. The strongest configuration, however, is \textit{ResNet-SSL frozen}, with a \SSLFrozenLosoOlder~ macro F1 score. Freezing the backbone preserves the full pretrained structure learned during UK Biobank pretraining rather than partially overwriting it during fine-tuning on a smaller task-specific dataset, \revised{consistent with the broader age diversity of UK Biobank providing a wider range of movement patterns}.

Across both protocols, richer representations progressively narrow the disparity between younger- and older-adult data. 
Under LOSO, performance for both populations improve along the ladder with only a modest 2\% decrease for younger adults, while the \revised{MyMove} older-adult curve rises more steeply, narrowing the gap from \textbf{\MlpLosoGap} to \textbf{\SSLFrozenLosoGap}. 
A similar pattern holds under zero-shot cross-dataset transfer: richer representations substantially improve older-adult performance from \MlpCdOlder~ to \SSLFrozenCdOlder~ while younger-adult performance stays stable or improves modestly; ResNet-SSL (frozen), at \SSLFrozenCdYoung, is only 4.1\% below the best with handcrafted features. The gap therefore narrows from \textbf{\MlpCdGap} to \textbf{\SSLFrozenCdGap}. 
\revised{To test whether SSL contributes to the observed gap narrowing, we exclude the handcrafted-feature baseline and group the remaining representations into configurations without versus with SSL (Figure~\ref{fig:mlp_head_loso_cd}(a,d)). A linear mixed-effects model of macro F1 with a cohort\,$\times$\,representation-type interaction and per-participant random intercepts shows that the effect of SSL differs by cohort under LOSO ($\chi^2(1){=}19.3$, $p<10^{-4}$; SSL boosts MyMove older adults by $14.4$ F1 points vs.\ $3.7$ for RealWorld younger) but not under cross-dataset ($\chi^2(1){=}0.55$, $p{=}0.46$).}

\begin{figure}[t]
    \centering
    \includegraphics[width=0.9\columnwidth]{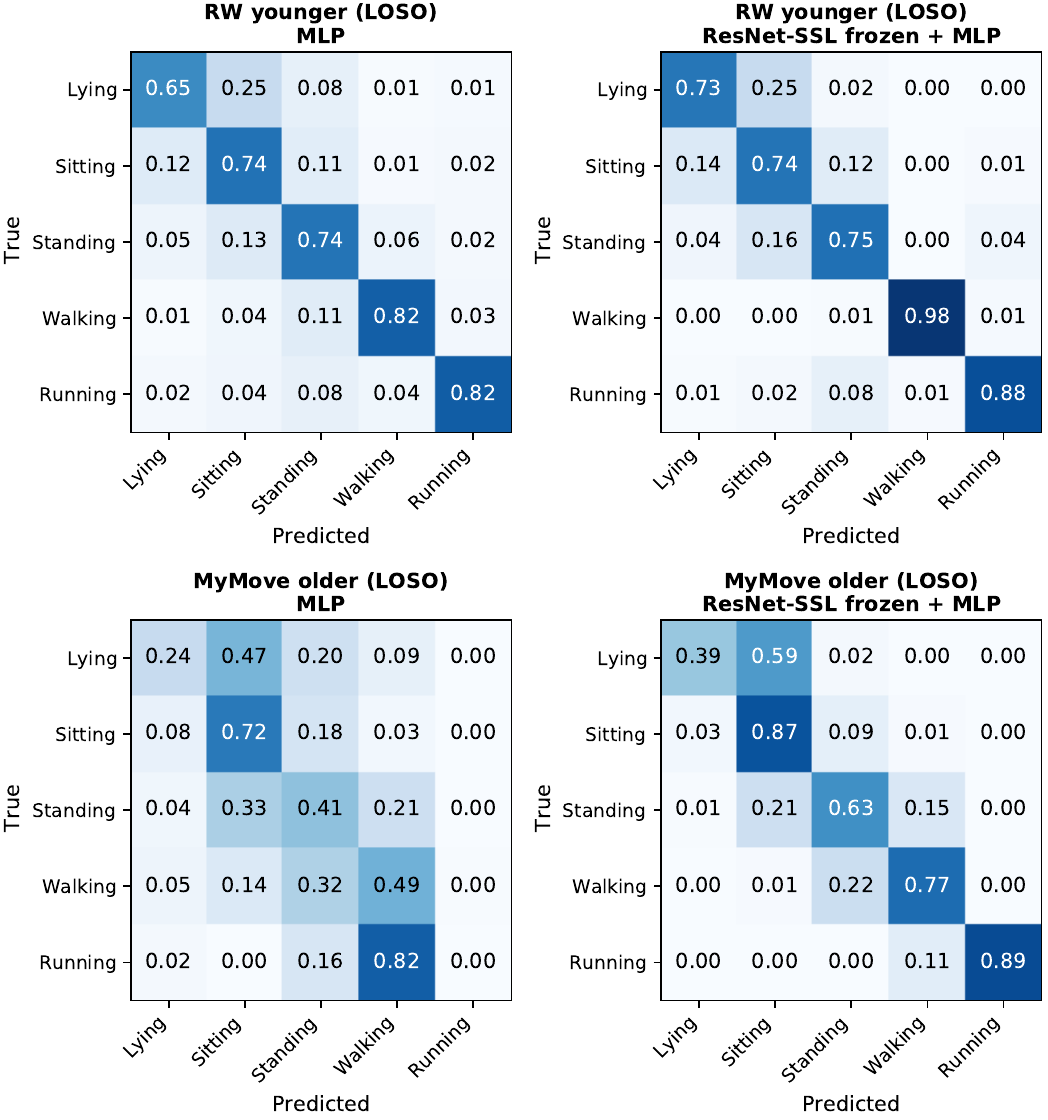}
        \vspace{-0.3cm}
    \caption{Adding the frozen SSL-pretrained backbone concentrates predictions more strongly along the diagonal, particularly for older adults, relative to the standalone MLP.}
    \label{fig:confusion_matrices}
    \Description[Four confusion matrices comparing MLP and frozen-SSL predictions for younger and older adults.]{Two-by-two grid of five-by-five confusion matrices, with true class on the vertical axis and predicted class on the horizontal axis over Lying, Sitting, Standing, Walking, and Running; darker cells mean higher probability and correct predictions lie on the diagonal. The top row is RealWorld younger adults, the bottom row is MyMove older adults; the left column is the standalone MLP and the right column adds a frozen self-supervised ResNet backbone. For younger adults the diagonal is already strong under the MLP and strengthens with the SSL backbone, for example Walking recall rising from 0.82 to 0.98. For MyMove older adults the MLP is weak and off-diagonal, with Lying recall only 0.24 (mostly confused with Sitting), Standing 0.41, and Walking 0.49; adding the frozen SSL backbone concentrates predictions along the diagonal, raising Lying to 0.39, Sitting to 0.87, Standing to 0.63, Walking to 0.77, and Running to 0.89. The largest gains appear for older adults, especially standing and walking, which the MLP frequently confused.}
        \vspace{-0.5cm}
\end{figure}

\begin{figure}[t]
    \centering
    \includegraphics[width=0.99\linewidth]{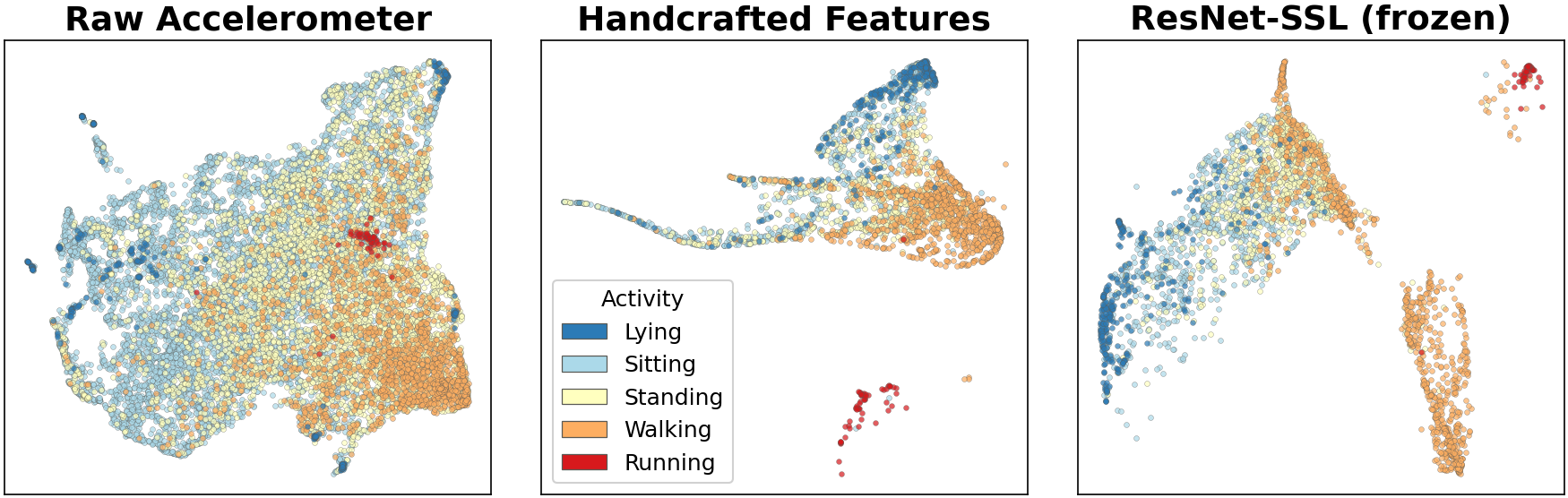}
        \vspace{-0.4cm}
    \caption{UMAP~\cite{mcinnes2018umap} projections for our older-adult dataset. Limited separation, \eg, for \textit{Standing} and \textit{Walking}, is shown.}
    \label{fig:umap_ssl_dataset}
    \Description[Three UMAP projections of older-adult windows under raw, handcrafted, and SSL representations.]{Three side-by-side UMAP scatter plots of the same MyMove older-adult accelerometer windows, colored by activity (Lying, Sitting, Standing, Walking, Running), under three representations: raw accelerometer signal, handcrafted features, and a frozen self-supervised ResNet. In the raw-signal panel the classes are heavily intermixed with no clear clusters. In the handcrafted-features panel a distinct Running cluster separates out, but Sitting, Standing, and Walking still overlap substantially. In the frozen self-supervised panel the structure is clearest: Running forms an isolated cluster, Lying and Sitting gather on one side and Walking on another, yet Standing remains partially overlapping with both Sitting and Walking. Richer representations increasingly separate the activities for older adults, but the stationary Standing-versus-Walking distinction stays hard to resolve even with the best representation.}
    \vspace{-0.5cm}
\end{figure}

Figure~\ref{fig:confusion_matrices} \revised{shows this trend qualitatively}. 
Relative to the standalone MLP, the \textit{ResNet-SSL frozen} concentrates predictions more strongly along the diagonal for both populations. For \revised{RealWorld} younger adults, the gains are most visible in edge activities, particularly \textit{Lying}, \textit{Walking}, and \textit{Running}, where SSL sharpens class separation and reduces spillover into neighboring classes. For \revised{MyMove} older adults the effect is broader: the SSL-pretrained representation improves separation across nearly all classes, substantially reducing the off-diagonal confusion of the standalone MLP. Almost all remaining misclassifications fall between neighboring classes, which is also inferable from the UMAP projections in Figure~\ref{fig:umap_ssl_dataset}.

\section{Discussion}
\label{sec:discussion}

Our findings suggest that progress in wearable HAR cannot be evaluated solely through aggregate benchmark performance. Across both LOSO and zero-shot cross-dataset evaluation, architectures achieving higher performance on younger-adult benchmarks did not improve proportionally on older-adult data, leaving a persistent performance gap.
In some cases, this gap widened. In contrast, richer representations, particularly those learned through self-supervised pretraining on large-scale age-diverse data, substantially improved older-adult performance while largely preserving younger-adult performance. 

These findings extend earlier observations of asymmetric transfer across age groups~\cite{delrosario2014comparison}, small but consistent performance declines with age~\cite{mardini2021age}, and improved HAR performance when older adults are explicitly included during training~\cite{ustad2023validation}. We add to this literature by showing that these patterns persist across multiple modern HAR architectures under both in-distribution and cross-dataset evaluation.
More broadly, our findings parallel observations in other machine-learning domains where benchmark progress fails to translate proportionally to deployment populations. This issue may be especially important in accessibility and aging applications, where benchmark populations often underrepresent the intended users. 
Cao et al.~\cite{cao2022different} reported a similar pattern in accessibility-oriented visual question answering: benchmark improvements did not translate proportionally to blind users, leaving a persistent performance gap that widened over time. In wearable HAR, evaluation on younger-adult benchmarks can therefore mask performance disparities while appearing to show steady progress.


\subsection{Limitations}
\revised{Our study does not isolate age as the sole source of the observed performance gap. Although wrist placement, sampling rate, and the activity taxonomy are matched across datasets, age remains entangled with differences in collection protocol, labeling methodology, and free-living versus scripted behavior. The RealWorld and MyMove age-outlier participants provide qualitative reference points but are insufficient to separate these factors conclusively.}

\revised{The semantic meaning of activity labels differs across populations. Controlled datasets typically record a scripted realization of each activity, whereas a free-living label in MyMove encompasses many behaviors. For MyMove older adults, \textit{Walking} spans leisure (27\%), dog walking (15\%), shopping (7.5\%), and a long tail of less frequent activities. By contrast, \textit{Walking} in our younger participants was almost entirely leisure walking (97\%). Consequently, part of the observed gap may reflect differences in the activities represented by shared labels, rather than differences in movement alone. Even a longitudinal study would not isolate age cleanly: as participants age, both activity repertoire and the semantic distribution of labels change, and over the same timescale wearable hardware, sensing technology, and machine-learning methods also evolve.}

\subsection{Interpreting the Gap}

\revised{To better understand the performance gap, we examined handcrafted features. Differences in feature means between age groups were comparable to, or smaller than, differences between datasets. Instead, the distinguishing factor was greater individual variability among older adults in orientation-invariant intensity (\textit{mag\_range}, \textit{rms}, \textit{mag\_std}) and cadence (\textit{f1}, \textit{f2}) features during \textit{Walking}, consistent with prior reports of increased gait variability with age~\cite{callisaya_ageing_2010,kang_separating_2008}. By contrast, younger adults exhibited greater variability in movement smoothness (\textit{jerk\_auto\_ratio}), anisotropy (\textit{cov\_condition}), and the PC1 zero-crossing interval (\textit{zc\_avg}). These observations suggest that the challenge lies less in systematic shifts of average movement than in greater diversity of movement patterns among older adults.}

This interpretation also aligns with our representation-learning results. Unlike increasingly complex end-to-end architectures, richer representations substantially reduced the performance gap, suggesting that broader movement variability is better addressed through transferable representations than through additional architectural capacity alone. Because labeled free-living older-adult datasets remain difficult to collect at scale~\cite{khayami2025verbal}, continued progress may depend increasingly on learning from large, age-diverse unlabeled datasets before adapting those representations to downstream HAR tasks. Continued self-supervised pretraining on older-adult recordings, domain adaptation, and test-time adaptation are therefore promising directions~\cite{tang2021selfhar,chang2020systematic,wang2024optimizationfree}.

More broadly, the remaining performance gap suggests that fully population-general HAR may remain challenging under current paradigms. The activities worth recognizing differ across individuals and life contexts~\cite{wang2024redefining}, and our findings indicate that even shared activity labels can represent different underlying behaviors. Rather than relying only on benchmark models, future wearable systems may benefit from combining broadly pretrained representations with personalized and teachable models that adapt to the activities and movement patterns most relevant to each user.
\begin{acks}
This work was supported by National Science Foundation (NSF) under Award \#1955568, and in part by NSF Award \#2229885; the National Institute on Disability, Independent Living, and Rehabilitation Research (NIDILRR Grant \#90REGE0024) funded by the Administration for Community Living (ACL), U.S. Department of Health and Human Services (HHS).
\end{acks}

\bibliographystyle{ACM-Reference-Format}
\bibliography{refs_full}

\end{document}